\documentclass{article}[12pt]
\begin{document}
\title{ Type-II seesaw mass models and baryon asymmetry}
\author{ Amal Kr. Sarma $^{a,b}$   , H. Zeen Devi$^{a}$ and N. Nimai Singh$^{a}${\footnote{Regular Associate of the Abdus Salam ICTP, Trieste, Italy}} }
\date{}
\maketitle
\begin{center}
$^{a}$Department of Physics, Gauhati University, Guwahati-781014, India.\\
$^{b}$Department of Physics, D. R. College, Golaghat-785621, India.\\
\end{center}
\begin{center}
\textbf{Abstract}
\end{center}
 We compute and also compare the contributions of canonical and noncanonical mass terms  towards baryon asymmetry by considering type-II seesaw mass models of neutrinos: Degenerate(3 varieties) , Normal hierarchical and Inverted hierarchical(2 varieties) . We have shown that for  particular choices of parameter `$\gamma$'( the so-called discriminator) for different neutrino mass models, the baryon asymmetry is largely dominated by canonical term. Within such   type-II seesaw scenario,  we find normal hierarchical (NHT3) mass model as the most  favourable choice of nature. 
\begin{flushleft}
\section{Introduction}
\end{flushleft}
The search for a suitable mass model consistent with the observations: masses, mixings and  baryon asymmetry, is a standing agenda of present day neutrino physics. With this motivation different neutrino mass models[1,2] were constructed on the basis of celebrated seesaw mechanism[3]. All the models were checked for masses, mixings and stability under radiative corrections. In the electroweak baryogenesis scenario[4,5], the heavy right-handed Majorana neutrinos  play a  decesive role. In this model of baryogenesis,  lepton asymmetry is produced by the decay of lightest of heavy Majorana neutrinos, and non-perturbative sphaleron[6] processes  convert this lepton asymmetry to baryon asymmetry. Again the heavy Majorana neutrinos are coupled to light left handed neutrinos via seesaw mechanism. In that sense any mass model constructed on the basis of seesaw mechanism should have the capacity to explain the observed baryon asymmetry.  Sakharov's three conditions[7] for baryogenesis: (1) baryon number violation, (2) C and CP violation and (3) out of equilibrium decay can be realised by the decay of  heavy right handed Majorana neutrinos. Recently a non-thermal scenario of baryogenesis for these models was considered[8], where our estimation[9] of baryon asymmetry for different mass models within the thermal leptogenesis scenario was extensively considered.
One point missed  in ref.(9) is that all the calculations to estimate the baryon asymmetry were confined to type-I seesaw term only, the contributions of non-canonical term were not considered. For completeness we compute the contribution of type-II term also. \\
We address in this paper, the    estimation of baryon asymmetry of the universe
 using the parameters which are  already fixed  at the seesaw stage within a common framework.
 This may possibly discriminate the  correct pattern of 
neutrino mass model in question. \\
In section 2  we   review the discrimination of neutrino mass models in type-II seesaw framework. In section 3, with a  brief discussion on   the expressions for lepton and baryon asymmetry, we present our  numerical calculations and results. Section 4   concludes  with a summary and discussion. The zeroth-order mass matrices for various models are collected in Appendix-A.
Important expressions related to $m^I_{LL}$ and $M_{RR}$ for three neutrino mass models 
are relegated to  Appendix-B.

\section{ Type-II seesaw mechanism and discrimination of neutrino mass models}

The type-I seesaw formula  relates the  light left-handed Majorana   neutrino mass matrix $m^I_{LL}$ and heavy right handed Majorana mass matrix $M_{RR}$ in a  simple way:
\begin{equation}
m_{LL}^I = -m_{LR}M_{RR}^{-1}m_{LR}^{T}.
\end{equation}
Here, $m_{LR}$ is the Dirac neutrino mass matrix.  In some L-R symmetric theories such as SO(10) GUT, the left-handed Higgs triplet $\Delta_L$ acquires  vacuum expectation value. Taking  this contribution, seesaw formula can be modified as[10]:
$m_{LL} = m_{LL}^{I} + m_{LL}^{II}$
where, the first term is the usual canonical seesaw term (1) and the second term  can be expressed as $m_{LL}^{II} = \gamma (M_W / v_R )^2 M_{RR}$. Such non-canonical seesaw formula (generally known as type-II seesaw formula) can be written as:
\begin{equation}
m_{LL} = -m_{LR}M_{RR}^{-1}m_{LR}^{T} + \gamma (M_W / v_R )^2 M_{RR}. 
\end{equation}
In the light of above type-II seesaw formula, the neutrino mass matrices $m_{LL}$ are realised for three different situations:
 (1) $m_{LL}^{II} >> m_{LL}^{I}$,  (2)  $m_{LL}^{II} \simeq m_{LL}^{I}$  and (3)
 $m_{LL}^{II} << m_{LL}^{I}$. It has been  reported that the  inequality  (1) can naturally provide a connection [11] between the large atmospheric mixing and $b-\tau$ unification in the context of the minimal supersymmetric SO(10) theory[12]. But we are interested to study the relative strength of these two contributions in all  three cases. The second term denoted by $m^{II}_{LL}$ is heavily constrained by the definition of $v_R$, and $v_R$ can be extracted from $M_{RR}$ appeared in $m^I_{LL}$, thus indicating the dependence of $m^{II}_{LL}$ on $m^I_{LL}$. This point is addressed in the present work.

The solar and atmospheric neutrino oscillation experiments usually measure only the mass-square differences and in general we have three possible patterns of neutrino masses[13]: (a) Degenerate $m_1\simeq m_2\simeq m_3$, (b) Inverted hierarchical $m_1\simeq m_2 >> m_3$ and (c) Normal hierarchical $m_1<< m_2 << m_3$. Depending upon the relative CP-phase of $(m_1 , m_2)$ pair, the degenerate and inverted hierarchical models   are further classified as: degenerate type-1A (DegT1A), degenerate type-1B (DegT1B), degenerate type-1C (DegT1C), inverted hierarchical Type-2A (InvT2A) and  inverted hierarchical Type-2B (InvT2B), and these
 zeroth-order left-handed Majorana mass matrices[2,13] are collected in Appendix A for ready reference. The present forms[1,2] of  various mass models given  in Appendix B, are derived from  seesaw mechanism(1).\\
         The vacuum expectation values $v_L$ and $v_R$ of left and right-handed Higgs triplets  are connected[14] to W-boson mass via $v_L v_R = \gamma M_W^2$   where  the free parameter $\gamma$ depends on  various couplings. Without fine tuning, we generally have  $\gamma\sim 1$, but we are searching the input values of $\gamma$ where the contribution of second term arising from the presence of left-handed Higgs  triplets  is just sufficient   to restore  the good predictions of $m^I_{LL}$ already acquired[2] in type-I seesaw framework. If  the second term is dominant,  the relevant good  predictions of $m^I_{LL}$ are  spoiled. This can be avoided with the proper choice of $\gamma$ values. We call this $\gamma$  parameter as `discriminator' of the models. A neutrino mass model is said to be favourable and hence stable when its canonical term dominates over the non-canonical(perturbative) term, and this condition is used here as a criterion for discriminating neutrino mass models[2].\\

  The Dirac neutrino mass matrix $m_{LR}$ appeared in seesaw formula can have any arbitrary   structure which either is   diagonal or non-diagonal, and it  plays a  significant role[1] in the construction of $m^I_{LL}$. In the seesaw mechanism, for a specific structure of $m^I_{LL}$, we can have three possible combinations of $m_{LR}$ and $M_{RR}$:
(a) both $m_{LR}$ and $M_{RR}$ are non-diagonal, (b) $m_{LR}$ diagonal and $M_{RR}$ non-diagonal,
(c) $m_{LR}$ non-diagonal and $M_{RR}$ diagonal.
These three combinations  can be realised in different physical situations. For example, when one calculate lepton asymmetry, one needs to consider the diagonal basis of heavy right-handed neutrinos, and combination (c) becomes  relevant. 
In the present calculations, the  diagonal form of $m_{LR}$ is chosen for different  neutrino  mass matrices.   To see this let us consider[2] the seesaw relation 
$m^I_{LL}= - m_{LR} M^{-1}_{R}m_{LR}^T$, where both $m_{LR}$ and $M_R$ are non-diagonal. 
Using some left and right handed rotations, the Dirac neutrino mass matrix can be diagonalised[15] as $m^{diag}_{LR} = U_L m_{LR} U^{\dag}_R$. In terms of diagonal basis of  $m_{LR}$, the seesaw relation reduces to  
\begin{equation}
m^I_{LL} = -U_L^{\dag} m_{LR}^{diag} M_{RR}^{-1} m_{LR}^{diag} U_L^*,
\end{equation}
where, $M^{-1}_{RR} = U_R M^{-1}_R U^T_R $.
It is assumed that eigenvalues of $m_{LR}^{diag}$ are hierarchical (similar to quarks or charged leptons). In absence of Dirac left handed rotations[15], we can set $U_L \sim 1$. For slight deviation from unity we can assume $U_L \simeq  U_{CKM}$, where $U_{CKM}$ is the  quark mixing matrix. Again this can be set to unity  as quarks mixings are very small. This type of approximations do not produce significant change in numerical calculations.  For $U_L \sim 1$ the eq.(3) reduces to $m^I_{LL} = -m_{LR}^{diag} M_{RR}^{-1} m_{LR}^{diag}$ where  $M_{RR}$ is in the diagonal basis of $m_{LR}$. We follow this representation in the present calculation.\\
In some Grand Unified Theory such as $SO(10)$ GUT , the possible structure[16] of $ m_{LR} = diag(\lambda^m,\lambda^n,1)v$ , where $v$ is the overall scale factor representing electroweak vacuum expectation values. In the  present calculation we take  $\lambda =0.3$ and $v =174 GeV$. We consider   two choices  of $(m,n)$ pair: case(i)  $(m,n)\equiv (6,2)$ for charged lepton  and (ii)  $(m,n)\equiv (8,4)$ for  up-quark mass matrices representing  the  Dirac neutrino mass matrix.\\

       As a representative  example, we consider  normal hierarchical mass model and the corresponding heavy right-handed neutrino mass matrix and light left-handed neutrino mass matrices  are[1,2] collected from Appendix B: 
\begin{flushleft}
\underline{Normal Hierarchical Type3(NHT3)}:
\end{flushleft}
\begin{center}
\begin{equation}
M_{RR}=\left(\begin{array}{ccc}
\lambda^{2m-1}&\lambda^{m+n-1}&\lambda^{m-1}\\
\lambda^{m+n-1}&\lambda^{m+n-2}&0\\
\lambda^{m-1}&0&1
\end{array}\right)v_{0},
\end{equation}
\end{center}
\begin{center}
\begin{equation}
-m_{LL}^{I}=\left(\begin{array}{ccc}
-\lambda^{4}&\lambda&\lambda^{3}\\
\lambda&1-\lambda&-1\\
\lambda^{3}&-1&1-\lambda^{3}
\end{array}\right)m_{0},
\end{equation}
\end{center}
where we take the input values $\lambda=0.3$, $m_0 = 0.03 eV$ and $v_0 = 1.01\times 10^{15} GeV$.

The numerical calculations are carried out for two choices of diagonal structure of Dirac neutrino mass matrix: case(i) Dirac neutrino mass matrix taken as charged lepton mass matrix and case(ii) Dirac neutrino mass matrix taken as up-quark mass matrix. Within the type-I seesaw framework, the left-handed neutrino mass matrix $m^I_{LL}$ leads to  correct neutrino mass parameters
 and mixing angles consistent with recent data[1,2]:
\begin{center}
$\Delta m^2_{21}= 9.04\times 10^{-5}eV^2, \Delta m^2_{23}= 3.01\times10^{-3}eV^2$\\
$\tan^2\theta_{12}= 0.55,\sin^22\theta_{23}= 0.98, \sin\theta_{13}= 0.074$.
\end{center}
In type-II seesaw scenario, one has to take care of  the non-canonical term $m^{II}_{LL}$. In order to maintain the existing good predictions of $m^I_{LL}$, one has to use the freedom in choosing $\gamma$. We make a search programme[2] for finding the value of the `minimum departure' of $\gamma$ from the canonical value of one , i.e.,$\gamma < 1.0$, in which the good prediction of neutrino masses and mixing parameters can again be restored in $m_{LL}$. In other words, such `least value' of $\gamma$ is just enough to suppress the perturbative effect due to type-II seesaw formula.  The value of $\gamma = 0.007$ is extracted for NHT3 mass model. The values of $\gamma$ in  case(i) and case(ii) for other neutrino mass models 
given in  Appendix B, are presented in Tables 2-3, along with  the corresponding predictions of neutrino oscillation parameters. Higher the values of $\gamma$, more the stability of the neutrino mass model in question.
Following the results of calculation in Tables 1-3. it has been observed that   InvT2A  model with $\gamma = 0.1$ gives most favourable choice of nature, followed by  NHT3 and InvT2B with $\gamma\sim 10^{-2}$ .
However, InvT2A model appears to be unstable under radiative corrections in MSSM. Now we are left with NHT3 and InvT2B as favourable candidates which are generally stable under radiative corrections in MSSM.\\
For further discrimination of the neutrino mass models, particularly  between NHT3 and InvT2B, we now consider baryon asymmetry estimation within the type-II seesaw formalism as an extra criteria  and this will be carried out  in the following section as the main thrust of the present invistigation. 

\section{Type-II seesaw mechanism and the  estimation of  baryon asymmetry}
 For our calculation of lepton asymmetry we consider the model[5] where the asymmetric decay  of the lightest of the heavy right-handed Majorana neutrinos, is assumed. 
In this model the physical  Majorana neutrino $N_{R}$ can decay into two modes:
\begin{center}
$N_{R}\rightarrow l_{L}+\overline{\phi}$\\

      $\rightarrow \overline{l}_{L}+\phi$
\end{center}
where $l_{L}$ is the lepton and $\bar{l}_{L}$ is the antilepton.
For CP-violating decay through the one-loop radiative correction by Higgs particle, the  branching ratio for these two decay modes is  likely to be different. 
The CP-asymmetry which is caused by the intereference of tree level with one-loop corrections
 for the decays of lightest of heavy right-handed Majorana neutrino $N_{1}$ is defined as[5,17]
\begin{center}
$\epsilon=\frac{\Gamma -\overline{\Gamma}}{\Gamma+\overline{\Gamma}}$
\end{center}
where, $\Gamma=\Gamma(N_{1}\rightarrow l_{L}\overline{\phi})$ and $\overline{\Gamma}=\Gamma(N_{1}\rightarrow \overline{l_{L}}\phi)$ are the decay rates.
Considering hierarchical structure of heavy Majorana neutrinos a  perturbative calculation from the interference between tree level and vertex plus self-energy diagrams, gives[18] the lepton asymmetry $\epsilon^I$ in  terms of light  neutrino mass matrix  $m_{LL}^I$  as :
\begin{equation}
\epsilon^I =  -\frac{3 M_1}{16\pi v^2}\frac{Im[(h^\ast m^I_{LL}h^{\dag})_{11}]}{(hh^{\dag})_{11}}.
\end{equation}
This accounts for the  type-I contribution. 
Again due to  the presence of Higgs triplet as a virtual particle in the decay of $N_1$,  we have the additional contribution [18] (type-II contribution) to lepton asymmetry  $\epsilon^{II}$  as:
\begin{equation}
\epsilon^{II} =  -\frac{3 M_1}{16 \pi v^2}\frac{Im[(h^\ast m^{II}_{LL}h^{\dag})_{11}]}{(hh^{\dag})_{11}}.
\end{equation}
For  quasi-degenerate spectrum i.e., for $M_1\simeq M_2<M_3$  the asymmetry  is largely affected by a resonance factor [19,20]  $R=|M_1|/{2(|M_2|-|M_1|)} $.
In such situation , the lepton asymmetry is modified [19,20] to:
\begin{center}
\begin{equation}
\epsilon=-\frac{M_2}{8 \pi v^2}\frac{Im[(h^{\ast} m_{LL} h^{\dag})_{11}]}{(hh^{\dag})_{11}}R
\end{equation}
\end{center}
The total asymmetry $\epsilon_1$ is the sum of $\epsilon^{I}$ and $\epsilon^{II}$.

 The sphaleron process convert[4,21] this lepton  asymmetry to baryon asymmetry. The baryon asymmetry $Y_B^{SM}$  is expressed in terms of the dilution factor $\kappa_1$  and lepton asymmetry $\epsilon_1$  as[22]
\begin{center}
\begin{equation}
Y_{B}^{SM}\equiv(\frac{n_{B}}{n_{\gamma}})^{SM}\simeq-1.08\times 10^{-2}\kappa_{1}\epsilon_{1}.
\end{equation}
\end{center}

For our calculation of baryon asymmetry we use the above  expression.

%%%%%%%%%%%%%%%%%%%%%%%%%%%%%%%%%%%%%%%%%%%%%%%%%%%%%%%%%%%%%%%%%%%%%%%%%%%%%%%%
How much the produced asymmetry is washed out is described by Boltzmann Equation and the  solution  can be described by a parameter  $\kappa_1$ known as dilution factor[14,23]:
\begin{center}
\begin{equation}
\kappa_1\simeq\left\{\begin{array}{ll}
\frac{0.3}{K(lnK)^{0.6}} &\textrm{if $10\le K\le10^{6}$}\\
\frac{1}{2\sqrt{K^2+9}} & \textrm{if $0\le K\le10$}\\
\end{array}\right.
\end{equation}
\end{center}
Where  $K=\tilde{m_{1}}/ m^\ast$  with  $\tilde{m_{1}}=\frac{(hh^{\dag})_{11}v^2}{M_{1}}$ is  the effective neutrino mass , $v$ the electroweak scale, $M_1$ the mass of $N_1$ ,  $h$ the matrix for the neutrino Yukawa couplings.  And $ m^\ast=\frac{16\pi^\frac{5}{2}}{3\sqrt{5}}g_\ast^\frac{1}{2}\frac{v^{2}}{M_{pl}}\simeq 1.08 \times10^{-3}eV$ is known as equilibrium neutrino mass[24] with $g^\ast=106.75$, $M_{pl}=1.2\times 10^{19}$ GeV.
Thus, for $\tilde{m_{1}}<m^\ast$,the heavy right-handed neutrino will satisfy the out-of-equilibrium
 condition. For our calculation we consider this condition also.
%%%%%%%%%%%%%%%%%%%%%%%%%%%%%%%%%%%%%%%%%%%%%%%%%%%%%%%%%%%%%%%%%%%%%%%%%%%%%%%%%%%%%%%%%%%%%%%%%%%%
\begin{flushleft}
\subsection{Numerical Calculation and results}
\end{flushleft}
 For our calculation of lepton asymmetry , we choose a basis $U_R$ where $M_{RR}^{diag}=U_R^T M_{RR}U_R=diag(M_1, M_2, M_3)$  with real and positive eigenvalues[15,25].We transform $m_{LR}=diag(\lambda^m, \lambda^n, 1)v$ to the $U_R$ basis by $m_{LR}\rightarrow m'_{LR} U_R$. In the prime basis the Yukawa coupling becomes $h = (m_{LR}U_R)/v$.
        For demonstration as an example,  we  consider here  the normal hierarchical mass model(NHT3) described in section 2.  
The corresponding heavy right-handed neutrino mass matrix and light left-handed neutrino mass matrices  are given in  eq.(4) and eq.(5). 

      For case(i): charged lepton mass matrix, $m_{LR}=m_{E}=diag(\lambda^{6}, \lambda^{2},1)v$, 
$(m,n)=(6,2)$,   $v_{0} = 1.01\times10^{15}$ GeV,  $m_{0}$= 0.03eV, $\gamma =0.007$.   
For the above  input values  the three heavy masses are   $M_{RR}^{diag}=diag(4.28\times10^9,1.16\times10^{10},3.84\times10^{13})$ in GeV.  From the construction of `$h$', we have  $(hh^{\dag})_{11}=5.31\times10^{-7}$  ,   $Im(h^* m_{LL}^I h^{\dagger})_{11}=2.45\times10^{-18}$,   $Im(h^* m_{LL}^{II} h^{\dagger})_{11}=2.52\times10^{-24}$,   $K=0.23$ and  $\kappa_1=0.17$ . These leads to lepton asymmetry   $\epsilon^I=5.92\times10^{-7}$ and  $\epsilon^{II}=6.09\times10^{-13}$  resulting  $\epsilon_1= \epsilon^I +\epsilon^{II}=5.92\times10^{-7}$.  The corresponding baryon asymmetry  is  $Y_B =-1.08\times 10^{-2} \kappa_1 \epsilon_1 =1.08\times10^{-9} $.

 For all other models including the normal hierarchy case(ii) we follow the same procedure to calculate the lepton  and baryon asymmetry. For case (ii) the Dirac neutrino mass matrix is taken as the up quark mass matrix i.e.,$m_{LR}=diag(\lambda^8,\lambda^4,1)v$. As  quasi-degenerate pattern of heavy masses appears  in Inverted hierarchy type2B (InvT2B), so for this model we consider the resonance enhancement factor $R$, and a different expression[19] $Y_B =-2.16\times 10^{-2} \kappa_1 \epsilon_1  $ is used for the estimation of baryon asymmetry. The estimated values of lepton and  baryon  asymmetry for all the models in Appendix B  are collected in tables 6-8.
As emphasised earlier, we use the only parameters fixed at seesaw stage in section 2  for estimating baryon asymmetry as an additional criteria for discriminating neutrino mass models. 

\section{Summary and Discussion}
 To summarise, we have considered  various neutrino mass models: degenerate models (DegT1A, DegT1B, DegT1C) , inverted hierarchical (InvT2A, InvT2B) and  normal hierarchical( NHT3). We have  estimated the contributions of type-I (canonical) and type-II ( non-canonical) seesaw mass term towards baryogenesis. Following the predictions of $SO(10)$ grand unified theory, we estimate the baryon asymmetry for two choices of Dirac neutrino mass matrix: case(i) Dirac neutrino mass matrix taken as charged lepton mass matrix and case(ii) Dirac neutrino mass matrix taken as up quark mass matrix.  For specific values of the discriminator  `$\gamma$' extracted from type-II seesaw formula[2], we observe that the baryon asymmetry is dominated by type-I term only. 

All the eigenvalues of $M_{RR}$ are collected in Table-4, and the mass parameter  $\tilde{m_1}$,  dilution factor $\kappa_1$  are presented in Table-5.  From  Table-5 it is clear that only inverted hierarchical model type2B (InvT2B)  and normal hierarchical model  type-3 (NHT3) 
satisfy the out-of-equilibrium decay condition. In these two models, the effective neutrino mass is less than the equilibrium neutrino mass  $m^{\ast}(\simeq1.08 \times10^{-3})$ i.e.,  $\tilde{m_{1}}<m^\ast$.\\
In Tables 6-8 we present the estimations of lepton and baryon asymmetry.
We present the numerical predictions of the neutrino mass-squared differences and three mixing angles in type-I and type-II seesaw framework[2]  in Tables 1-3. Only two models  i.e., inverted hierarchical model  type2A (InvT2A) and  normal hierarchical  model  type 3 (NRT3) satisfy all the neutrino oscillation parameters (Tables 1-3). However, the inverted hierarchical Type-2A(InvT2A) model is not stable under radiative corrections in MSSM, whereas normal hierarchical type 3 (NHT3) is stable.

If we consider the observed baryon asymmetry[26] $Y_B=(6.1^{-0.3}_{-0.2})\times 10^{-10}$ , a  competitive nature appears for both Degenerate type-1 model (DegT1A) and Normal hierarchical type3(NHT3). But considering the mixing parameters  and  stability condition, one can easily rule out the degenerate type 1A (DegT1A) model.
 For NHT3  the mass of $M_{1}$ is also within  the Ibarra-Davidson  bound [27]. Considering the above results we have found that in type-II scenario   normal hierarchical type 3( NHT3)  model is the most  favourable  choice of nature. 

The present calculation may also be useful to discriminate  the choices of Dirac neutrino mass matrix. The charged lepton, up-quark and down-quark approximation  of Dirac neutrino mass matrix may not be the right choice of nature. Once a neutrino mass  model is fixed by experiment, then  one can search for the possible structure of Dirac neutrino mass matrix by constraining observed baryon asymmetry to such model[28].  
\pagebreak

Table-1: Predicted values of the  solar and atmospheric neutrino mass-squared differences  and three mixing parameters 
calculated from $m_{LL}^{I}$  derived from type-I seesaw formula  in the Appendix-B.
\begin{center}
\begin{tabular}{cccccc}\hline
Type&$\Delta m^{2}_{21}[10^{-5}eV^{2}]$&$\Delta m^{2}_{23}[10^{-3}eV^{2}]$&$\tan^{2}\theta_{12}$&$\sin^{2}2\theta_{23}$&$\sin\theta_{13}$\\
\hline
DegT1A&$8.80$&$2.83$&$0.98$&$1.0$&$0.0$\\
DegT1B&$7.91$&$2.50$&$0.27$&$1.0$&$0.0$\\
DegT1C&$7.91$&$2.50$&$0.27$&$1.0$&$0.0$\\
InvT2A&$8.36$&$2.50$&$0.44$&$1.0$&$0.0$\\
InvT2B&$9.30$&$2.50$&$0.98$&$1.0$&$0.0$\\
NHT3&$9.04$&$3.01$&$0.55$&$0.98$&$0.074$\\
\hline
\end{tabular}
\end{center}

Table-2:Predicted values of the solar and atmospheric neutrino mass-squared 
differences and three mixing parameters along with the values of $\gamma$,  extracted from $m_{LL}$ in type-II seesaw formula for case (i). 
\begin{center}
\begin{tabular}{ccccccc}\hline
Type & $\gamma$ & $\bigtriangleup m^{2}_{21}[10^{-5}eV^{2}]$ & 
 $\bigtriangleup m^{2}_{23}[10^{-3}eV^{2}]$ & $\tan^{2}\theta_{12}$
 & $\sin^{2}2\theta_{23}$ & $\sin\theta_{13}$ \\
\hline
 DegT1A & $10^{-5}$ & $8.45$ & $2.73$ & $0.98$ & $1.00$ & $0.0$\\

  DegT1B & $10^{-4}$ & $7.97$ & $2.30$ & $0.28$ & $1.00$ & $0.0$\\

 DegT1CC & $10^{-5}$ & $7.93$ & $2.47$ & $0.27$ & $1.00$ & $0.0$\\

 InvT2A & $0.1$ & $8.20$ & $2.50$ & $0.49$ & $1.00$ & $0.0$\\

 InvT2B & $0.009$ & $9.40$ & $2.40$ & $0.98$ & $1.00$ & $0.01$\\

 NHT3 & $0.007$ & $9.41$ & $2.98$ & $0.54$ & $0.98$ & $0.09$\\
\hline
\end{tabular}
\end{center}

Table-3:Predicted values of solar and atmospheric neutrino mass-squared differences, and three mixing
parameters along with the values of $\gamma$, extracted from $m_{LL}$ in type-II seesaw formula  for case (ii).

\begin{center}
\begin{tabular}{ccccccc}\hline
Type & $\gamma$ & $\bigtriangleup m^{2}_{21}[10^{-5}eV^{2}]$ & 
 $\bigtriangleup m^{2}_{23}[10^{-3}eV^{2}]$ & $\tan^{2}\theta_{12}$
 & $\sin^{2}2\theta_{23}$ & $\sin\theta_{13}$      \\ \hline

 DegT1A & $10^{-5}$ & $8.56$ & $2.74$ & $0.98$ & $1.00$ & $0.0$\\

 DegT1B & $10^{-4}$ & $7.69$ & $2.30$ & $0.27$ & $1.00$ & $0.0$\\

 DegT1CC & $10^{-5}$ & $7.69$ & $2.54$ & $0.29$ & $1.00$ & $0.0$\\

 InvT2A & $0.1$ & $8.3$ & $2.5$ & $0.47$ & $1.00$ & $0.0$\\

 InvT2B & $0.02$ & $9.40$& $2.40$ & $0.98$ & $1.00$ & $0.0$\\

 NHT3 & $0.007$ & $9.18$ & $2.80$ & $0.55$ & $0.98$ & $0.07$\\

\hline
\end{tabular}
\end{center}

\pagebreak

Table-4: The three right-handed Majorana neutrino masses in GeV for both case (i) and case (ii). The expressions of 
$M_{RR}$ are collected from Appendix-B.
\begin{center}
\begin{tabular}{cc|c}
\hline
Type   &  Case(i):$|M_{j}|$            &  Case(ii):$|M_{j}|$\\           
\hline
DegT1A & $4.28\times 10^9,1.16\times10^{10},3.84\times10^{13}$  & $3.47\times10^7,9.42\times10^7,3.81\times10^{13}$\\
DegT1B & $4.05\times10^7,6.16\times10^{11},7.60\times^{13}$  & $3.28\times10^5,4.98\times10^9,7.60\times10^{13}$\\
DegT1C & $4.05\times10^7,6.69\times10^{12},6.69\times10^{12}$  & $3.28\times10^5,4.85\times10^{11},7.81\times10^{11}$\\
InvT2A & $3.28\times10^{8},9.70\times10^{12},6.79\times 10^{16}$  & $2.64\times10^6,7.92\times10^{10},6.70\times10^{16}$\\
InvT2B & $5.6527\times10^{10},5.6532\times10^{10},5.38\times10^{16}$  & $4.5971\times10^8,4.5974\times10^8,5.34\times10^{16}$\\
NHT3  & $6.51\times10^{10},7.97\times10^{11},1.01\times10^{15}$  & $5.27\times10^8,6.45\times10^9,1.01\times10^{15}$\\
\hline
\end{tabular}
\end{center}

Table-5 :Estimation of  the  values of effective mass parameter $\tilde{m_{1}}$ in eV, and dilution factor $\kappa_1$.
\begin{center}
\begin{tabular}{cccc|ccc}
\hline
 & For & case(i) & & For & case(ii)&\\
\hline
Type   &$\tilde{m}_1$ &$ K$ &$ \kappa_1$ & $\tilde{m}_1$ &$ K$ &$ \kappa_1$ \\             \hline
DegT1A & $3.76\times10^{-3}$&$3.48 $&$ 0.11$ & $3.74\times10^{-3}$&$3.46$&$0.11$\\
DegT1B & $0.40$&$370$& $2.79\times 10^{-4}$& $0.40$&$370$&$ 2.79\times 10^{-4}$\\
DegT1C & $0.40$&$370$&$2.79\times 10^{-4}$& $0.40$&$370$&$ 2.79\times 10^{-4}$\\
InvT2A & $0.05$&$46.3$&$ 2.9\times 10^{-3}$& $0.05$&$46.3$&$2.9\times 10^{-3}$\\
InvT2B & $2.85\times10^{-4}$&$0.26$&$ 0.17$& $2.83\times10^{-4}$&$0.26$&$0.17$\\
NHT3   & $2.47\times10^{-4}$&$0.23$&$0.17$ & $2.47\times10^{-4}$&$0.23$&$0.17$\\
\hline
\end{tabular}
\end{center}

Table 6: Calculation of lepton asymmetry $\epsilon^I$ and $\epsilon^{II}$ for case(i) for respective neutrino mass models given in  Appendix-B.
\begin{center}
\begin{tabular}{ccc|c}
\hline
Type  &  $\epsilon^I$     & $\epsilon^{II}$ & $\epsilon_1 = \epsilon^I + \epsilon^{II}$\\
\hline
DegT1A & $2.10\times10^{-6}$  & $1.19\times10^{-15}$ & $2.10\times10^{-6}$\\
DegT1B & $2.66\times10^{-18}$ & $3.88\times10^{-27}$ & $2.66\times10^{-18}$\\
DegT1C & $1.74\times10^{-18}$ & $2.24\times10^{-25}$ & $1.74\times10^{-18}$ \\
InvT2A & $1.59\times10^{-14}$ & $5.30\times10^{-26}$ & $1.59\times10^{-14}$\\
InvT2B & $1.47\times10^{-2}$  & $2.52\times10^{-13}$  & $1.47\times10^{-2}$\\
NRT3   & $5.92\times10^{-7}$  & $6.09\times10^{-13}$  & $5.92\times10^{-7}$\\
\hline
\end{tabular}
\end{center}
\pagebreak
Table 7: Calculation of lepton asymmetry $\epsilon^I$ and $\epsilon^{II}$ for case(ii) for respective neutrino mass models given in  Appendix-B.
\begin{center}
\begin{tabular}{ccc|c}
\hline
Type  &  $\epsilon^I$     & $\epsilon^{II}$ & $\epsilon_1 = \epsilon^I + \epsilon^{II}$\\
\hline
DegT1A & $1.71\times10^{-8}$  & $7.98\times10^{-20}$ & $1.17\times10^{-8}$\\
DegT1B & $2.16\times10^{-20}$ & $2.54\times10^{-31}$ & $2.16\times10^{-20}$\\
DegT1C & $1.69\times10^{-20}$ & $1.43\times10^{-26}$ & $1.69\times10^{-20}$ \\
InvT2A & $1.27\times10^{-16}$ & $2.46\times10^{-28}$ & $1.27\times10^{-16}$\\
InvT2B & $1.62\times10^{-4}$  & $5.08\times10^{-17}$ & $1.62\times10^{-4}$\\
NRT3   & $4.78\times10^{-9}$  & $4.90\times10^{-17}$ & $4.78\times10^{-9}$\\
\hline
\end{tabular}
\end{center}

Table-8: Calculation of   baryon asymmetry $Y_{B}$  for respective neutrino mass models given in Appendix-B.
\begin{center}
\begin{tabular}{c c|c}
\hline
Type   &  $Y_{B}^{SM} case(i)$ & $Y_{B}^{SM} case(ii)$\\                       
\hline
DegT1A & $2.49\times10^{-9}$   & $2.03\times10^{-11}$\\
DegT1B & $8.00\times10^{-24}$  & $6.50\times10^{-26}$ \\
DegT1C & $5.20\times10^{-24}$  & $5.10\times10^{-26}$ \\
InvT2A & $4.97\times10^{-19}$  & $3.98\times10^{-21}$ \\
InvT2B & $5.40\times10^{-5}$   & $5.94\times10^{-7}$  \\
NHT3   & $1.08\times10^{-9}$   & $8.80\times10^{-12}$\\
\hline
\end{tabular}
\end{center}
\pagebreak
%%%%%%%%%%%%%%%%%%%%%%%%%%%%%%%%%%%%%%%%%%%%%%%%%%%%%%%%%%%%%%%%%%%%
\section*{Appendix A}

We list here the  zeroth-order left-handed Majorana neutrino mass 
matrices[2,13] with texture zeros, $m_{LL}^I$, corresponding to three models of neutrinos, viz., 
degenerate (DegT1A, DegT1B, DegT1C), inverted hierarchical (InvT2A, InvT2B) and normal hierarchical (NHT3).

\begin{center}
\begin{tabular}{ccc}\hline
Type  & $m_{LL}$ & $m_{LL}^{diag}$\\ \hline \\
\ [DegT1A]    &${ \left(\begin{array}{ccc}
  0 & \frac{1}{\sqrt{2}} & \frac{1}{\sqrt{2}}\\ 
 \frac{1}{\sqrt{2}} & \frac{1}{2} & -\frac{1}{2}\\
 \frac{1}{\sqrt{2}} & -\frac{1}{2} & \frac{1}{2} 
\end{array}\right)}m_{0}$ & $Diag(1,-1,1)m_{0}$\\

\\
\ [DegT1B]    &${ \left(\begin{array}{ccc}
  1 & 0 & 0\\ 
 0 & 1 & 0\\
 0 & 0 & 1 
\end{array}\right)}m_{0}$ & $Diag(1,1,1)m_{0}$\\
\\

\ [DegT1C]    &${ \left(\begin{array}{ccc}
  1 & 0 & 0\\ 
 0 & 0 & 1\\
 0 & 1 & 0 
\end{array}\right)}m_{0}$ & $Diag(1,1,-1)m_{0}$\\ \hline
\\
\ [InvT2A]    &${ \left(\begin{array}{ccc}
  1 & 0 & 0\\ 
 0 & \frac{1}{2} & \frac{1}{2}\\
 0 & \frac{1}{2} & \frac{1}{2} 
\end{array}\right)}m_{0}$ & $Diag(1,1,0)m_{0}$\\
\\

\ [InvT2B]    &${ \left(\begin{array}{ccc}
  0 & 1 & 1\\ 
 1 & 0 & 0\\
 1 & 0 & 0 
\end{array}\right)}m_{0}$ & $Diag(1,-1,0)m_{0}$\\ \hline
\\

\ [NHT3]    &${ \left(\begin{array}{ccc}
  0 & 0 & 0\\ 
 0 & \frac{1}{2} & -\frac{1}{2}\\
 0 & -\frac{1}{2} & \frac{1}{2} 
\end{array}\right)}m_{0}$ & $Diag(0,0,1)m_{0}$  \\ 

\\ \hline
\end{tabular}
\end{center}
\pagebreak
\section*{Appendix B}

Here we collect[1,2]  the various right handed Majorana mass matrices $M_{RR}$ and the light left handed neutrino mass matrix $m^I_{LL}$ along with the input values of the parameters.
\begin{flushleft}
\underline{Degenerate Type1A(DegT1A)}:
\end{flushleft}
\begin{center}
\begin{displaymath}
\mathbf{M_{RR}}=\left(\begin{array}{ccc}
 -2\delta_{2}\lambda^{2m}&(\frac{1}{\sqrt{2}}+\delta_{1})\lambda^{m+n}&(\frac{1}{\sqrt{2}}+\delta_{1})\lambda^{m}\\
(\frac{1}{\sqrt{2}}+\delta_{1})\lambda^{m+n}&(1/2+\delta_{1}-\delta_{2})\lambda^{2n}&(-1/2+\delta_{1}-\delta_{2})\lambda^{n}\\
(\frac{1}{\sqrt{2}}+\delta_{1})\lambda^{m}&(-1/2+\delta_{1}-\delta_{2})\lambda^{n}&(1/2+\delta_{1}-\delta_{2})
\end{array}\right)v_{0}
\end{displaymath}
\end{center}
\begin{center}
\begin{displaymath}
\mathbf{-m_{LL}^{I}}=\left(\begin{array}{ccc}
 (-2\delta_{1}+2\delta_{2})&(\frac{1}{\sqrt{2}}-\delta_{1})&(\frac{1}{\sqrt{2}}-\delta_{1})\\
(\frac{1}{\sqrt{2}}-\delta_{1})&(1/2+\delta_{2})&(-1/2+\delta_{2})\\
(\frac{1}{\sqrt{2}}-\delta_{1})&(-1/2+\delta_{2})&(1/2+\delta_{2})
\end{array}\right)m_{0}
\end{displaymath}
\end{center}
Here, $\delta_{1}=0.0061875 ,\delta_{2}=0.0031625 ,\lambda=0.3, m_{0}=0.4$eV and $v_0=7.57\times10^{13}GeV$,  $\gamma = 10^{-5}$.
\begin{flushleft}
\underline{Degenerate Type1B(DegT1B)}:
\end{flushleft}
\begin{center}
\begin{displaymath}
\mathbf{M_{RR}}=\left(\begin{array}{ccc}
(1+2\delta_{1}+2\delta_{2})\lambda^{2m}&\delta_{1}\lambda^{m+n}&\delta_{1}\lambda^{m}\\
\delta_{1}\lambda^{m+n}&(1+\delta_{2}\lambda^{2n}&\delta_{2}\lambda^{n}\\\delta_{1}\lambda^{m}&\delta_{2}\lambda^{n}&(1+\delta_{2})
\end{array}\right)v_{0}
\end{displaymath}
\end{center}
\begin{center}
\begin{displaymath}
\mathbf{-m_{LL}^{I}}=\left(\begin{array}{ccc}
(1-2\delta_{1}-2\delta_{2})&-\delta_{1}&-\delta_{1}\\
-\delta_{1}&(1-\delta_{2})&-\delta_{2}\\
-\delta_{1}&-\delta_{2}&(1-\delta_{2})
\end{array}\right)m_{0}
\end{displaymath}
\end{center}
Here, $\delta_{1}=7.2\times10^{-5},\delta_{2}=3.9\times10^{-3},\lambda=0.3,m_{0}=0.4$eV and $v_0=7.57\times10^{13}GeV$, $\gamma = 10^{-4}$.

\begin{flushleft}
\underline{Degenerate Type1C(DegT1C)}:
\end{flushleft}
\begin{center}
\begin{displaymath}
\mathbf{M_{RR}}=\left(\begin{array}{ccc}
(1+2\delta_{1}+2\delta_{2})\lambda^{2m}&\delta_{1}\lambda^{m+n}&\delta_{1}\lambda^{m}\\
\delta_{1}\lambda^{m+n}&\delta_{2}\lambda^{2n}&(1+\delta_{2})\lambda^{n}\\\delta_{1}\lambda^{m}&(1+\delta_{2})\lambda^{n}&\delta_{2}
\end{array}\right)v_{0}
\end{displaymath}
\end{center}
\begin{center}
\begin{displaymath}
\mathbf{-m_{LL}^{I}}=\left(\begin{array}{ccc}
(1-2\delta_{1}-2\delta_{2})&-\delta_{1}&-\delta_{1}\\
-\delta_{1}&-\delta_{2}&(1-\delta_{2})\\
-\delta_{1}&(1-\delta_{2})&-\delta_{2}
\end{array}\right)m_{0}
\end{displaymath}
\end{center}
Here, $\delta_{1}=7.2\times10^{-5},\delta_{2}=3.9\times10^{-3},\lambda=0.3,m_{0}=0.4$eV and $v_0=7.57\times10^{13}GeV$, $\gamma = 10^{-5}$.
\begin{flushleft}
\underline{Inverted Hierarchical Type2A(InvT2A)}:
\end{flushleft}
\begin{center}
\begin{displaymath}
\mathbf{M_{RR}}=\left(\begin{array}{ccc}
\eta(1+2\epsilon)\lambda^{2m}&\eta\epsilon\lambda^{m+n}&\eta\epsilon\lambda^{m}\\
\eta\epsilon\lambda^{m+n}&1/2\lambda^{2n}&-(1/2-\eta)\lambda^{n}\\
\eta\epsilon\lambda^{m}&-(1/2-\eta)\lambda^{n}&1/2
\end{array}\right)\frac{v_{0}}{\eta}
\end{displaymath}
\end{center}
\begin{center}
\begin{displaymath}
\mathbf{-m_{LL}^{I}}=\left(\begin{array}{ccc}
(1-2\epsilon)&-\epsilon&-\epsilon\\
-\epsilon&\frac{1}{2}&(\frac{1}{2}-\eta)\\
-\epsilon&(\frac{1}{2}-\eta)&\frac{1}{2}
\end{array}\right)m_{0}
\end{displaymath}
\end{center}
Here, $\eta=0.0045 ,\epsilon=0.0055,\lambda=0.3 ,m_{0}=0.05$eV and $v_0=6.06\times10^{13}GeV$, $\gamma = 0.1$.
\begin{flushleft}
\underline{Inverted Hierarchical Type2B(InvT2B)}:
\end{flushleft}
\begin{center}
\begin{displaymath}
\mathbf{M_{RR}}=\left(\begin{array}{ccc}
\lambda^{2m+7}&\lambda^{m+n+4}&\lambda^{m+4}\\
\lambda^{m+n+4}&\lambda^{2n}&-\lambda^{n}\\
\lambda^{m+4}&-\lambda^{n}&1
\end{array}\right)v_{0}
\end{displaymath}
\end{center}
\begin{center}
\begin{displaymath}
\mathbf{-m_{LL}^{I}}=\left(\begin{array}{ccc}
0&1&1\\
1&-(\lambda^{3}-\lambda^{4})/2&-(\lambda^{3}+\lambda^{4})/2\\
1&-(\lambda^{3}+\lambda^{4})/2&-(\lambda^{3}-\lambda^{4})/2
\end{array}\right)m_{0}
\end{displaymath}
\end{center}
Here,$\lambda=0.3 , m_{0}=0.035$eV and $v_0=5.34\times10^{16}GeV$. For case(i) $\gamma = 0.009$ and for case(ii) $\gamma = 0.02$.

\begin{flushleft}
\underline{Normal Hierarchical Type3(NHT3)}:
\end{flushleft}
\begin{center}
\begin{displaymath}
\mathbf{M_{RR}}=\left(\begin{array}{ccc}
\lambda^{2m-1}&\lambda^{m+n-1}&\lambda^{m-1}\\
\lambda^{m+n-1}&\lambda^{m+n-2}&0\\
\lambda^{m-1}&0&1
\end{array}\right)v_{0}
\end{displaymath}
\end{center}
\begin{center}
\begin{displaymath}
\mathbf{-m_{LL}^{I}}=\left(\begin{array}{ccc}
-\lambda^{4}&\lambda&\lambda^{3}\\
\lambda&1-\lambda&-1\\
\lambda^{3}&-1&1-\lambda^{3}
\end{array}\right)m_{0}
\end{displaymath}
\end{center}
Here,$\lambda=0.3,m_{0}=0.03$eV and $v_0=1.01\times10^{15}GeV$, $\gamma = 0.007$.
\section*{Acknowledgements}
One of us (AKS) would like to thank  UGC(NER), India,  for awarding a fellowship under FIP programme, Xth plan.

\end{document}